\documentclass[preprint,showpacs,preprintnumbers,amsmath,amssymb]{revtex4-1}
 \usepackage{graphicx}
 \usepackage{dcolumn}
 \usepackage{bm}
 \usepackage{ulem}
 \usepackage{enumitem}
\def\q{\bm{q}}
\def\p{\bm{p}}
\def\k{\bm{k}}

\usepackage[usenames]{color}

\renewcommand\sout{\bgroup \color{red} \ULdepth=-.5ex \ULset}

\begin{document}

\title{Quarkonium dissociation in perturbative QCD}
\author{Juhee Hong}
\affiliation{Department of Physics and Institute of Physics and Applied Physics, Yonsei University,
Seoul 03722, Korea}
\author{Su Houng Lee}
\affiliation{Department of Physics and Institute of Physics and Applied Physics, Yonsei University,
Seoul 03722, Korea}
\date{\today}

\begin{abstract}
For weakly bound quarkonia, we rederive the next-to-leading order cross 
sections of quarkonium dissociation by partons that include the hard thermal 
loop (HTL) resummation. 
Our results calculated with an effective vertex from the Bethe-Salpeter 
amplitude  reduce to those obtained by potential nonrelativistic QCD 
(pNRQCD) in the relevant kinematical limit, and they can be used in a wide 
temperature range applicable to heavy quark systems in heavy ion collisions. 
Based on the lattice computation of the temperature-dependent binding energy, 
our numerical analysis on $\Upsilon(1S)$ indicates that 
at high temperature the dominant mechanism for quarkonium dissociation is 
inelastic parton scattering as expected in the quasifree approximation, 
while it is the gluo-dissociation at low temperature. 
By comparing with the momentum diffusion coefficient of a heavy quark, we 
discuss possible $O(g)$ corrections to the next-to-leading order thermal width. 
\end{abstract}

\maketitle

\section{Introduction}

Quarkonia are an important probe of high temperature and density matter 
produced in relativistic heavy ion collisions. 
The suppression of quarkonia and their sequential melting provide information 
about the formation of quark-gluon plasma, the thermal properties of 
the matter, and the heavy quark potential at finite temperature. 
Originally, color screening in deconfined quark-gluon plasma has 
been thought to prevent heavy quarks to form a bound state \cite{satz}. 
Now quarkonia are believed to dissolve at high temperature 
not because the binding energy vanishes but because the thermal width becomes 
as large as the reduced binding energy. 

Quarkonium suppression has been investigated for a long time, but 
their yields still need to be understood quantitatively. 
For recent reviews, see Refs. \cite{rapp-review,strickland-review}. 
In addition to color-screening, quarkonia yields are affected by other 
mechanisms including Landau damping, feed-down, initial state conditions, 
cold nuclear matter effects, and possible regenerations near the threshold. 
In this work, we focus on the dissociation mechanisms of quarkonia. 
Especially, we are interested in the ground state of bottomonium 
$\Upsilon(1S)$, since it survives at high temperature up to $\sim 600$ MeV 
\cite{strickland-review}.  
Recently, bottomonium suppression has been observed in Pb+Pb collisions at 
$\sqrt{s_{NN}}=2.76$ TeV by the CMS collaboration \cite{cms} and 
in U+U collisions at $\sqrt{s_{NN}}=193$ GeV by the STAR collaboration 
\cite{star}. 

From a partonic picture,  
there are two main mechanisms of quarkonium dissociation.  
The first is gluo-dissociation, $g+\Upsilon\rightarrow Q+\bar{Q}$, 
where $\Upsilon$ is quarkonium and 
$Q(\bar{Q})$  a heavy (anti)quark. 
Gluo-dissociation, also known as the thermal breakup of the heavy 
quark-antiquark color singlet state in effective field theory, 
breaks the bound state by absorbing a gluon from thermal medium.  
The dipole interaction of color charge with a gluon \cite{peskin,peskin-bhanot} 
has been used to study dissolution of quarkonia.    
The second mechanism is inelastic parton scattering, 
$p+\Upsilon\rightarrow p+Q+\bar{Q}$ with $p=g,q,\bar{q}$. 
This is related to the Landau damping phenomenon which results from 
scattering of hard particles in thermal bath exchanging spacelike gluons. 
Quarkonium dissociation by inelastic parton scattering has been investigated 
at the quasifree approximation, where the reaction is taken care of by the 
sum of $p+Q\rightarrow p+Q$ and $p+\bar{Q}\rightarrow p+\bar{Q}$ \cite{rapp}.
For a tightly bound state near the phase transition, 
gluo-dissociation is an efficient process \cite{peskin}, while 
inelastic parton scattering is expected to be dominant 
when loosely bound quarkonia scatter with hard particles 
\cite{rapp,rapp-review}. 
In an effective field theory framework inelastic parton scattering is also 
expected to be dominant at high temperature, but the quasifree approximation 
might overestimate the dissociation cross sections \cite{Song:2010ix,pnrqcd-nlo}. 
For this reason, we need to calculate inelastic parton scattering exactly and 
compare its contribution with that of gluo-dissociation.

Recently, there have been rigorous and formal developments in the calculations 
on these issues. 
The quark-antiquark static potential has been derived in finite 
temperature QCD, where it has been found that the potential develops 
an imaginary part that results in the thermal width \cite{width}. 
The real part of the potential is a screened Coulomb potential, and 
the imaginary part is induced by the Landau damping. 
Furthermore, in an effective field theory framework another part of 
the imaginary potential has been found out to arise by the singlet-to-octet 
thermal transition \cite{pnrqcd-prd}. 
In different temperature regimes at weak coupling, 
potential nonrelativistic QCD (pNRQCD) has been used to 
study quarkonium dissociation 
\cite{pnrqcd-prd,pnrqcd-melt,pnrqcd-lo,pnrqcd-nlo}. 

While the low-energy effective field theory is an important formal development, 
the various results valid at different energy and temperature scales are 
problematic when trying to use them in realistic estimates for the fate of 
quarkonium, as the hierarchy of scales varies during the time evolution in 
relativistic heavy ion collisions. 
Moreover, since these imaginary parts of the potential are related to the 
two mechanisms of quarkonium dissociation, we might be able to obtain the 
resulting thermal width from the perspective of scattering processes 
($g+\Upsilon\rightarrow Q+\bar{Q}$ and 
$p+\Upsilon\rightarrow p+Q+\bar{Q}$).
Therefore, our aim is to introduce the partonic picture for quarkonium 
dissociation that reduces to the formal limit of the effective field theory 
calculations at the relevant kinematical regime, but that can also interpolate 
between different temperature scales which are applicable to heavy quark 
systems in evolving plasmas.

This work is organized as follows. 
In Sec. \ref{gluo-diss}, we discuss gluo-dissociation up to next-to-leading 
order. 
In Sec. \ref{inelastic}, we use an effective vertex derived from the leading 
order dissociation to calculate inelastic parton scattering which contributes 
to dissolution at next-to-leading order. 
In Sec. \ref{numeric}, we present a numerical analysis for weakly coupled 
$\Upsilon(1S)$ based on the lattice data of the temperature-dependent binding 
energy. 
In Sec. \ref{g-corr}, we discuss possible $O(g)$ corrections to the 
next-to-leading order quarkonium dissociation. 
Finally, we summarize our results in Sec. \ref{summary}.

\section{Gluo-dissociation}
\label{gluo-diss}

\begin{figure}

\includegraphics[width=0.35\textwidth]{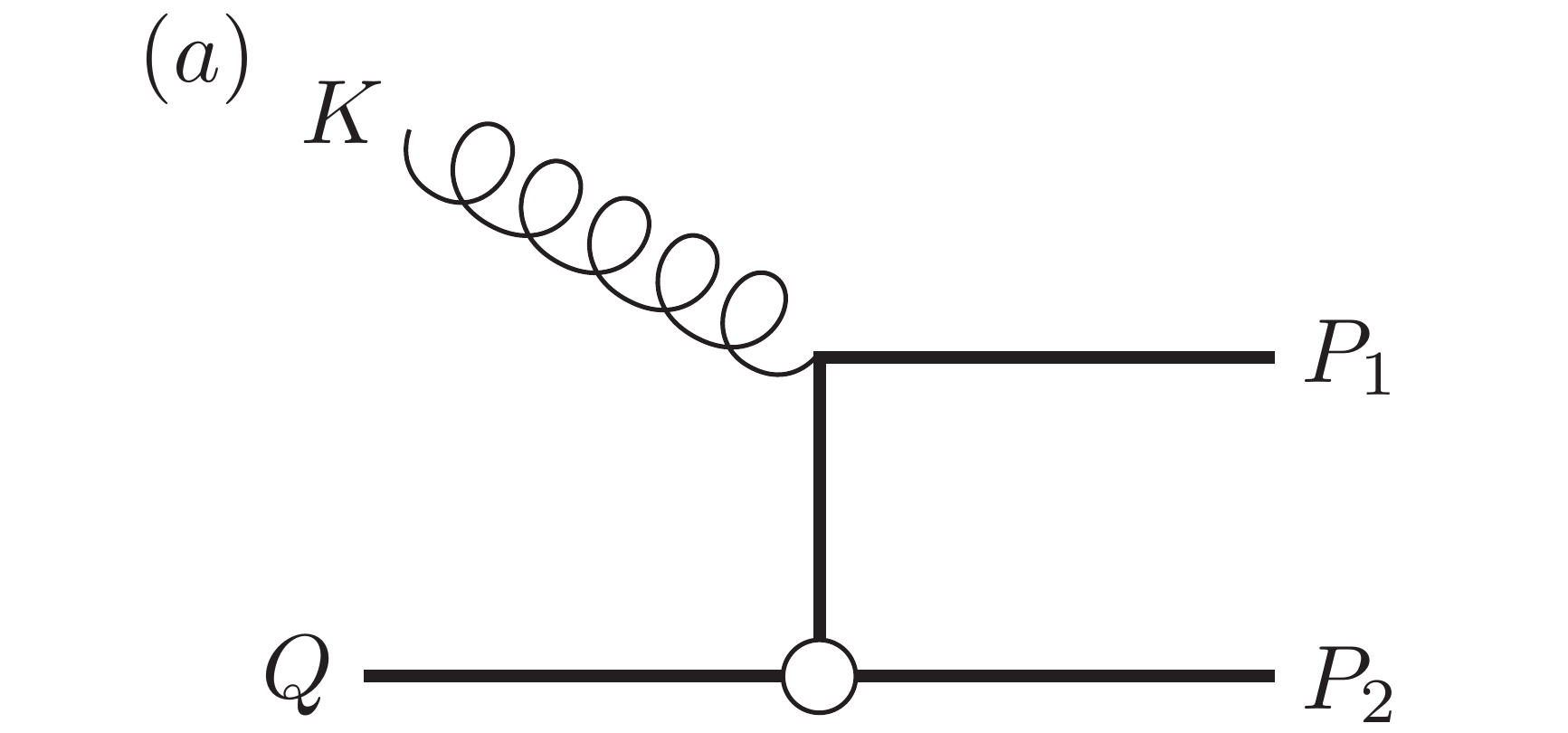}
\qquad\qquad
\includegraphics[width=0.33\textwidth]{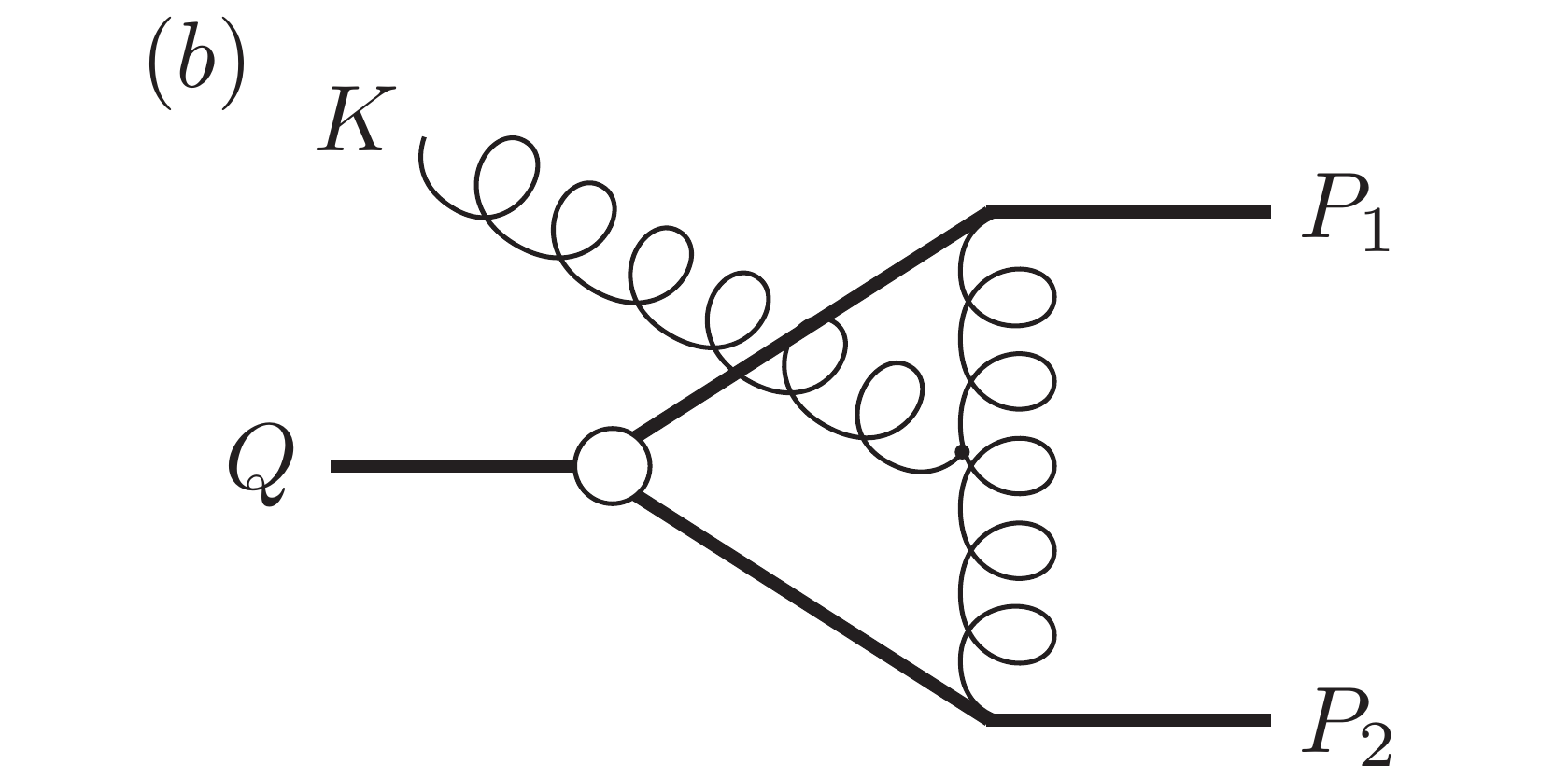}
\caption{The leading order gluon-quarkonium interactions. 
Thick solid lines denote quarkonium ($Q$) or heavy (anti)quarks 
($P_1$, $P_2$), and wiggly lines are gluons.
}
\label{lo-process}
\end{figure}

In the heavy quark limit, heavy quark systems can be treated by using 
perturbative QCD with the interquark potential reducing to the Coulomb type.
In Ref. \cite{peskin}, Peskin has calculated the interactions between partons 
and heavy quark bound states by performing the operator 
product expansion on gluon insertions in the heavy quark states.
As an application, the gluo-dissociation cross section of quarkonium has been
obtained at leading order \cite{peskin-bhanot}, and later the same cross 
section has been rederived by using the Bethe-Salpeter amplitude \cite{lee-lo}. 
In $1/N_c$ expansion in the large $N_c$ limit, the heavy quark-antiquark pair 
after dissociation is noninteracting such that the dissolution is induced by 
the processes shown in Fig. \ref{lo-process} (and another similar to (a) but 
$P_1,P_2$ exchanged).

For $\q\sim\p_1,\p_2\gg\k$, the sum of the leading processes gives the 
following scattering amplitude \cite{lee-lo}:
\begin{equation}
\mathcal{M}_{LO}^{\mu\nu}=-g\sqrt{\frac{m_\Upsilon}{N_c}}\left[\k\cdot\frac{\partial
\psi(\p)}{\partial\p}\delta^{\mu 0}+k_0\frac{\partial\psi(\p)}{\partial p^i}
\delta^{\mu i}\right]\delta^{\nu j}\bar{u}(P_1)\frac{1+\gamma_0}{2}\gamma^j
\frac{1-\gamma^0}{2}T^av(P_2) \, ,
\label{mtx-lo}
\end{equation}
where $\psi(\p)$ is the normalized wave function for a bound state with the 
relative momentum, $\p=(\p_1-\p_2)/2$. 
The matrix element squared is
\begin{equation}
|\mathcal{M}|_{LO}^2=\frac{8(N_c^2-1)}{N_c}g^2m^2m_\Upsilon k_0^2 \, \vert
\nabla\psi(\p)\vert^2 \, ,
\end{equation}
where we have used the transverse polarization 
($\delta^{ij}-\hat{k}^i\hat{k}^j$) for gluon. 
Dividing by the quarkonium polarization $d_\Upsilon$ and 
the gluon degeneracy $d_g=2(N_c^2-1)$, 
the averaged scattering cross section is given by 
\begin{multline}
\label{sig-lo}
\sigma_{LO}(k)=\frac{1}{4\sqrt{(Q\cdot K)^2-m_\Upsilon^2m_k^2}}
\int\frac{d^3\p_1}{(2\pi)^32p_{10}}
\int\frac{d^3\p_2}{(2\pi)^32p_{20}}
\\
\times
(2\pi)^4\delta^4(Q+K-P_1-P_2)
\frac{1}{d_\Upsilon d_g}|\mathcal{M}|^2_{LO} \, .
\end{multline} 

In the heavy quark limit, the heavy quark-antiquark pair is assumed to be 
weakly bound so that the leading order ground state wave function of 
quarkonium can be described by the Coulombic bound state \cite{peskin} 
\begin{equation}
\label{coulomb}
|\nabla\psi_{1S}(\p)|=2^5\sqrt{\pi}\frac{a_0^{7/2}\p}{[(a_0\p)^2+1]^3} \, ,
\end{equation}
where $a_0=16\pi/(N_cmg^2)$ is the Bohr radius. 
In the rest frame of the plasma where quarkonium is approximately at rest, 
we obtain \cite{lee-lo,peskin-bhanot} 
\begin{equation}
\label{crossLO}
\sigma_{LO}(k)=\frac{2^7g^2a_0^2}{d_\Upsilon N_c}\frac{E^{7/2}(k-E)^{3/2}}{k^5} \, .
\end{equation}
Here $E=2m-m_{\Upsilon}>0$ is the binding energy of quarkonium, 
and the relation $a_0^2=1/(mE)$ is satisfied for the Coulombic binding energy. 

To obtain the cross section at leading order, we have assumed that the 
external gluon is massless and $k\sim k_0\gg m_D$. 
On the other hand, at finite temperature gluons acquire the thermal 
self-energy that leads to thermal mass $\sim m_D$. 
To accommodate such self-energy effects, we note that the phase space of the 
initial gluon involves the following on-shell condition for the dispersion 
relation at leading order: 
\begin{equation}
\label{dispersion-lo}
1=\int dk_0 \, \delta(k_0-k)=\int dk_0 \, 2k_0 \, \delta(k_0^2-k^2) \, .
\end{equation}
In the case of $k\ll T$ and $k_0^2-k^2\sim m_D^2$, 
the hard thermal loop (HTL) resummation is needed and the dispersion relation 
for transverse gluons is then given by 
\begin{equation}
\label{dispersion-nlo}
k_0^2-k^2-{\rm{Re}}\, \Pi_T(k_0,k)=0 \, .
\end{equation}
In the Coulomb gauge, the HTL self-energies are \cite{weldon} 
\begin{eqnarray}
\label{HTLselfe}
\Pi_L(k_0,k)&=&m_D^2\left[1-\frac{k_0}{2k}\log\left(\frac{k_0+k}{k_0-k}\right)
\right] \, ,
\nonumber\\ 
\Pi_T(k_0,k)&=&\frac{m_D^2}{2}\left[\frac{k_0^2}{k^2}-
\frac{(k_0^2-k^2)k_0}{2k^3}\log\left(\frac{k_0+k}{k_0-k}\right)\right] \, , 
\end{eqnarray}
which do not have imaginary parts for time-like ($k_0^2> k^2$) gluon.  
Therefore,  
at next-to-leading order Eq. (\ref{dispersion-lo}) is extended as follows: 
\begin{multline}
\int dk_0 \, 
2k_0 \, \delta\left(k_0^2-k^2-\frac{m_D^2}{2}\left[\frac{k_0^2}{k^2}-
\frac{(k_0^2-k^2)k_0}{2k^3}\log\left(\frac{k_0+k}{k_0-k}\right)\right]\right)
\\
\simeq
\left[1+\frac{m_D^2}{4k^2}\left(\log\left(\frac{8k^2}{m_D^2}
\right)-2\right)\right]^{-1} 
\, ,
\label{dispersionT}
\end{multline}
where we have used $k_0\simeq k$ and in the argument of the logarithm  
$k_0-k\simeq \frac{m_D^2}{4k}$.  

By involving Eq. (\ref{dispersionT}) in the leading order result, 
we obtain the next-to-leading order cross section for gluo-dissociation, 
\begin{equation}
\sigma_{NLO}(k)\simeq
\left[1-\frac{m_D^2}{4k^2}\left(\log\left(\frac{8k^2}{m_D^2}
\right)-2\right)\right]\sigma_{LO}(k) \, ,
\label{gluo-nlo}
\end{equation}
where the effect of the relative velocity between quarkonium and gluon is 
included here through the definition of the thermal width given in Eq. 
(\ref{gluo-gamma}). 
If we neglect the final state rescattering effects of the 
unbound heavy quark-antiquark pair in the large $N_c$ limit, 
Eq. (\ref{gluo-nlo}) agrees with the pNRQCD results 
\cite{pnrqcd-lo,pnrqcd-nlo} 
for the $mv\gg T\gg E\gg m_D$ case (in this regime, gluo-dissociation is the 
dominant dissociation mechanism in the effective field theory framework).

With a thermal distribution function of an incoming gluon, 
the thermal width of gluo-dissociation can be calculated as \cite{rapp} 
\begin{equation}
\label{gluo-gamma}
\Gamma=d_g\int \frac{d^3\k}{(2\pi)^3} \, \sigma(k) \, n(k) \, . 
\end{equation} 
To break a bound state, the magnitude of an incoming gluon momentum is 
required to be larger than the binding energy, $|\k|\geq E$.

\section{Inelastic parton scattering}
\label{inelastic}

In this section, we introduce an effective 
vertex which is derived from the leading order scattering processes of 
Fig. \ref{lo-process}. 
In terms of the vertex, we calculate 
inelastic parton scatterings which contribute to quarkonium dissociation 
at next-to-leading order.

\begin{figure}
\includegraphics[width=0.45\textwidth]{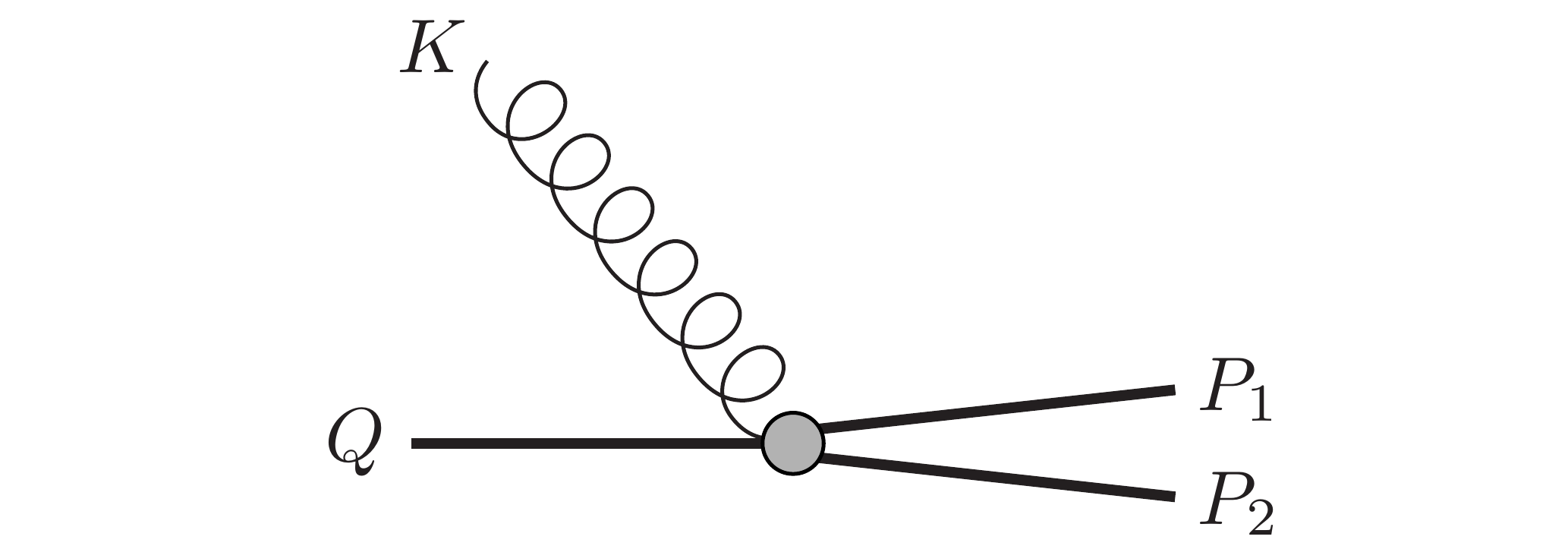}
\caption{The effective vertex derived from the sum of the leading 
processes shown in Fig. \ref{lo-process}. 
}
 \label{vertex}
\end{figure}

From the leading order scattering amplitude of Eq. (\ref{mtx-lo}), 
the following effective vertex can be derived  
\cite{lee-nlo}:  
\begin{equation}
V^{\mu\nu}(K)=-g\sqrt{\frac{m_\Upsilon}{N_c}}\left[\k\cdot\frac{\partial
\psi(\p)}{\partial\p}\delta^{\mu 0}+\left(\frac{\p^2}{m}+E\right)
\frac{\partial\psi(\p)}{\partial p^i}
\delta^{\mu i}\right]\delta^{\nu j}\frac{1+\gamma^0}{2}\gamma^j 
\frac{1-\gamma^0}{2}T^a \, ,
\label{eff-vertex}
\end{equation}
and  denoted by Fig. \ref{vertex}. 
With the energy conservation, $\p^2/m+E\simeq k_0$ ($k_0$ is the energy 
transfer), this vertex is similar to the potential nonrelativistic QCD 
(pNRQCD) vertex.  
The difference is that our approach involves $1\pm\gamma^0$ matrices because 
of the nonrelativistic treatment of heavy quarks, while in pNRQCD the octet 
propagator is employed to calculate the imaginary part of the potential. 
As will be shown below, our results agree with those by pNRQCD in a certain 
regime where inelastic parton scattering becomes dominant. 
The reason is that the rescattering effects in the octet propagator become 
subleading and the chromoelectric dipole interaction can be taken into 
account through the effective vertex of Eq. (\ref{eff-vertex}). 
This vertex reflects the dipole interaction of color charge 
for the leading gluon-quarkonium interaction 
considered in Refs. \cite{peskin,peskin-bhanot}.

For quarkonium dissolution by inelastic parton scattering, 
we consider barely bound quarkonia with near-threshold energy so that 
partons in heat bath separate a heavy quark-antiquark pair almost 
collinearly. 
In the quarkonium rest frame, the energy transfer is small 
($k_{10}\simeq k_{20}$) and 
the only contribution is from the longitudinal gluon part of the 
effective vertex in Eq. (\ref{eff-vertex}). 
This situation is similar to the momentum diffusion of a heavy quark where 
the internal gluon attached to the Wilson line is 
longitudinal \cite{sch-moore1,sch-moore2}.

\begin{figure}
\includegraphics[width=0.3\textwidth]{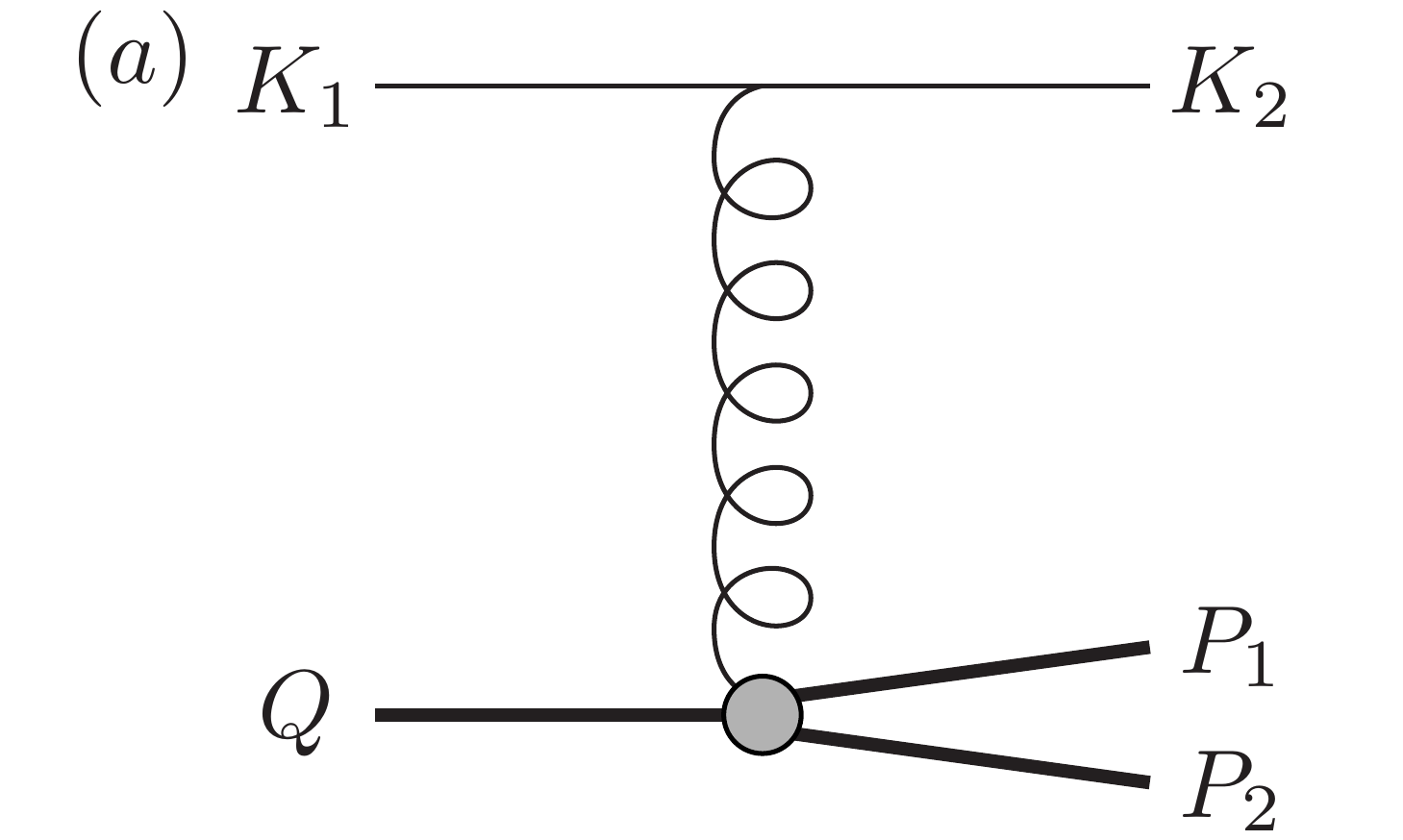}
\qquad\qquad\qquad
\includegraphics[width=0.3\textwidth]{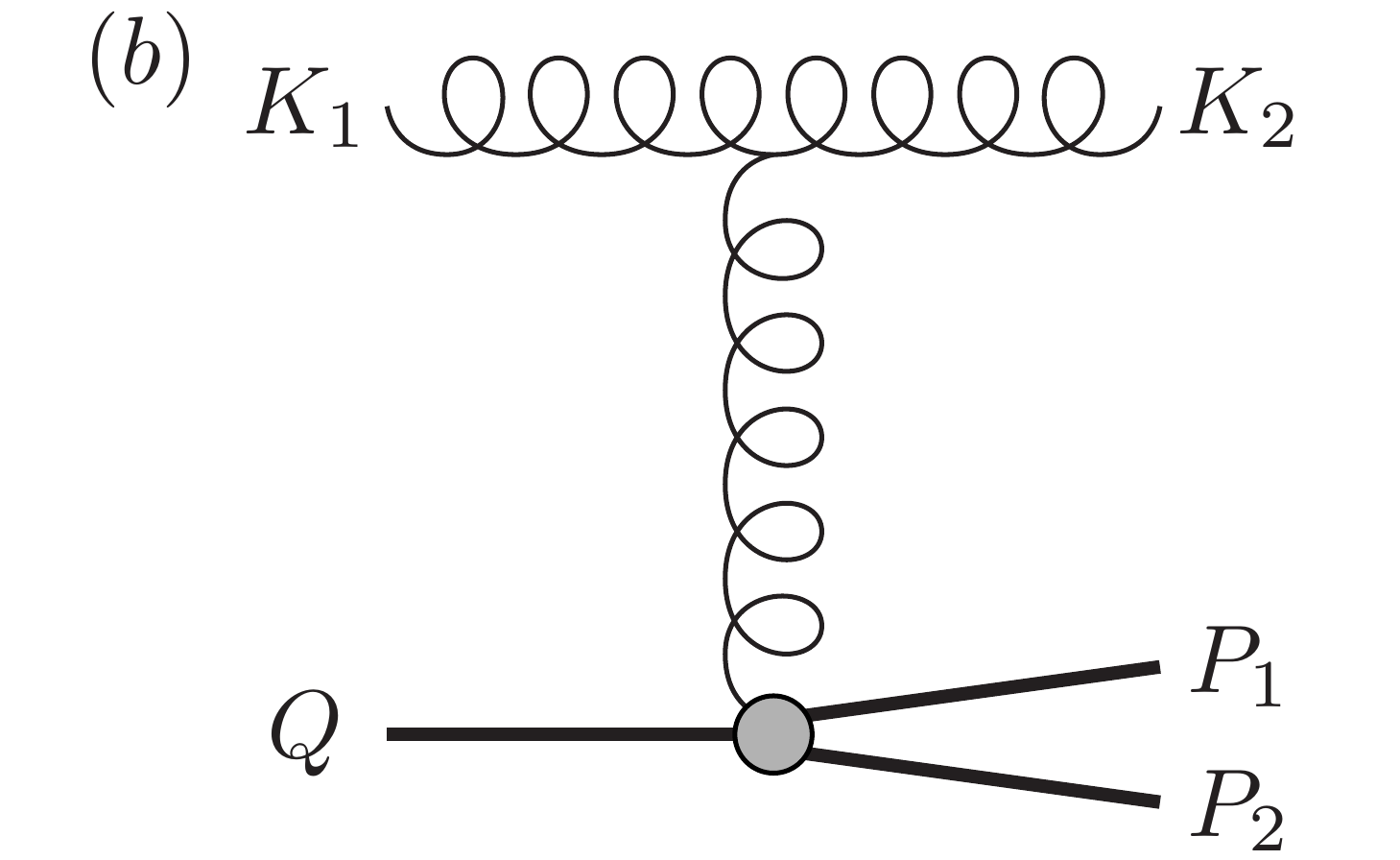}
\caption{The inelastic parton scattering processes which contribute to 
quarkonium dissociation at next-to-leading order (NLO). 
(b) gives O(g) corrections to the NLO results if all the gluons are soft.
}
 \label{NLOdiag}
\end{figure}

Inelastic parton scattering shown in Fig. \ref{NLOdiag} can be calculated 
by using the effective vertex. 
For $\k_1,\k_2\gg \k_1-\k_2$, 
the corresponding scattering amplitudes are given by  
\begin{eqnarray}
\mathcal{M}_{NLO}^{\mu(q)}&=&g
\bar{u}(K_2)\gamma_\nu T^au(K_1)\frac{1}{(K_1-K_2)^2-\Pi(K_1-K_2)}
\bar{u}(P_1)V^{\nu\mu}(K_1-K_2)v(P_2) \, ,
\nonumber\\
\mathcal{M}_{NLO}^{\mu\nu\lambda(g)}&=&-igf^{abc}
\bar{u}(P_1)V^{\rho\mu}(K_1-K_2)v(P_2) \frac{1}{(K_1-K_2)^2-\Pi(K_1-K_2)}
\nonumber\\
&&\quad\times
\left[g^\nu_\rho(2K_1-K_2)^\lambda-g_\rho^\lambda(K_1-2K_2)^\nu-g^{\lambda\nu}
(K_1+K_2)_\rho\right] \, .
\label{mtx-nlo}
\end{eqnarray}
In the Coulomb gauge, the HTL propagator of longitudinal gluon is 
$i/[k^2+\Pi_L(k_0,k)]\simeq i/(k^2+m_D^2)$ for small energy transfer. 
The matrix elements squared are then 
\begin{eqnarray}
\label{mtx2-nlo}
|\mathcal{M}|_{NLO}^{2(q,\bar{q})}&\simeq&
\frac{16N_f(N_c^2-1)}{N_c}g^4m^2m_\Upsilon\vert\nabla\psi(\p)\vert^2
\frac{(\k_{1}-\k_{2})^2k_{10}^2}{[(\k_1-\k_2)^2+m_D^2]^2}
\left[1+\frac{\k_1\cdot \k_2}{k_{1}k_{2}}\right] \, ,
\nonumber\\
|\mathcal{M}|_{NLO}^{2(g)}&\simeq&16(N_c^2-1)
g^4m^2m_\Upsilon\vert\nabla\psi(\p)\vert^2
\frac{(\k_{1}-\k_{2})^2k_{10}^2}{[(\k_1-\k_2)^2+m_D^2]^2}
\left[1+\frac{(\k_1\cdot\k_2)^2}{k_1^2k_2^2}\right] \, ,
\end{eqnarray}
where in the gluon-induced reaction we have used transverse polarizations 
for external gluons.  

By integrating over the phase space and dividing by the initial flux, 
the dissociation cross section is obtained as 
\begin{multline}
\sigma_{NLO}(k_1)=\frac{1}{4\sqrt{(Q\cdot K_1)^2-m_\Upsilon^2m_{k_1}^2}}
\int\frac{d^3\k_2}{(2\pi)^32k_{20}}
\int\frac{d^3\p_1}{(2\pi)^32p_{10}}
\int\frac{d^3\p_2}{(2\pi)^32p_{20}}
\\
\times
(2\pi)^4\delta^4(Q+K_1-K_2-P_1-P_2)
\frac{1}{d_\Upsilon d_p}|\mathcal{M}|_{NLO}^2 \, ,
\end{multline}
where $d_p$ ($p=g,q$) is the degeneracy factor of the incoming parton. 
To perform the phase space integration, we proceed as follows 
\cite{amy-llog,moore-teaney}. 
First, $\p_2$ integration is done by using the three-dimensional delta 
function, and $\k_2$ integration is changed to $\k=\k_1-\k_2$. 
For the (anti)quark-quarkonium scattering, we have 
\begin{multline}
\sigma_{NLO}^{(q,\bar{q})}(k_1)\simeq\frac{N_f(N_c^2-1)g^4m_\Upsilon k_{10}}{2(2\pi)^5N_cd_\Upsilon d_q|Q\cdot K_1|}
\int d^3\k \int d^3\p_1 \, 
\\
\times \,
\delta(q_0+k_{10}-k_{20}-p_{10}-p_{20}) \,
\vert\nabla\psi(\p)\vert^2
\frac{k^2}{(k^2+m_D^2)^2}
\left(2-\frac{k^2}{2k_{1}^2}\right) \, . 
\end{multline}
Second, we introduce a dummy variable $\omega$, 
\begin{equation}
\delta(q_0+k_{10}-k_{20}-p_{10}-p_{20})=
\int d\omega \, \delta(\omega+q_0-p_{10}-p_{20}) \, 
\delta(\omega-k_{10}+k_{20}) \, ,
\end{equation}
and the $\k$-$\p_1$ phase space is integrated over $k$, $p_1$, 
$\theta_{kk_1}$, $\theta_{kp_1}$, and $\phi_{k;k_1p_1}$, where 
$\theta_{kk_1}$($\theta_{kp_1}$) is the angle between two vectors $\k$ and 
$\k_1$($\p_1$) and $\phi_{k;k_1p_1}$ is the angle between the $\k$-$\k_1$ 
plane and the $\k$-$\p_1$ plane. 
For dissolution of weakly bound quarkonia with $\q\sim\p_1,\p_2\gg\k$, we use 
the following relations  
\begin{eqnarray}
\delta(\omega+q_0-p_{10}-p_{20})&\simeq&\frac{p_{10}}{kp_1}
\delta\left(\cos\theta_{kp_1}-\frac{\omega p_{10}}{kp_1}\right) \, ,
\nonumber\\
\delta(\omega-k_{10}+k_{20})&=&\frac{k_2}{kk_1}
\delta\left(\cos\theta_{kk_1}-\frac{\omega}{k}-\frac{k^2-\omega^2}{2kk_1}
\right) \, ,
\end{eqnarray}
to perform the integrations over two polar angles. 
Then the cross section becomes 
\begin{equation}
\label{phaseq}
\sigma_{NLO}^{(q,\bar{q})}(k_1)
\simeq
\frac{N_f(N_c^2-1)g^4m_\Upsilon k_{10}}{(2\pi)^3N_cd_\Upsilon d_q|Q\cdot K_1|}
\int_0^{2k_1} dk  \int dp_1 \, p_1^2 \, 
\vert\nabla\psi(\p)\vert^2
\frac{k^3}{(k^2+m_D^2)^2}
\left(2-\frac{k^2}{2k_{1}^2}\right) \, ,
\end{equation}
where $\omega$ has been integrated over the range $(-\frac{kp_1}{p_{10}},\frac{kp_1}{p_{10}})$. 
Finally, in the quarkonium rest frame ($\q=0$) $p_1$ integration is 
changed to $p^2$ integration for a bound state with $|\nabla\psi(\p)|^2$ as a 
function of $p^2$.   
After integrating over $p^2$ and $k$, we obtain 
\begin{eqnarray}
\label{nlo-q}
\sigma_{NLO}^{(q,\bar{q})}(k_1)
&\simeq&\frac{3N_f(N_c^2-1)g^4a_0^2}{4N_cd_\Upsilon d_q\pi}\left[\log\left(\frac{4k_{1}^2}{m_D^2}\right)-2
\right] \, . 
\end{eqnarray}
Here, we have used the Coulombic bound state Eq. (\ref{coulomb}), but 
the relation $a_0^2=1/(mE)$ need not be satisfied in general. 
Similarly, the gluon-quarkonium scattering cross section is determined as 
\begin{equation}
\label{nlo-g}
\sigma_{NLO}^{(g)}(k_1)\simeq\frac{3(N_c^2-1)g^4a_0^2}{4d_\Upsilon d_g\pi}\left[\log\left(\frac{4k_{1}^2}{m_D^2}\right)-2
\right] \, .
\end{equation}

\begin{figure}
\includegraphics[width=0.45\textwidth]{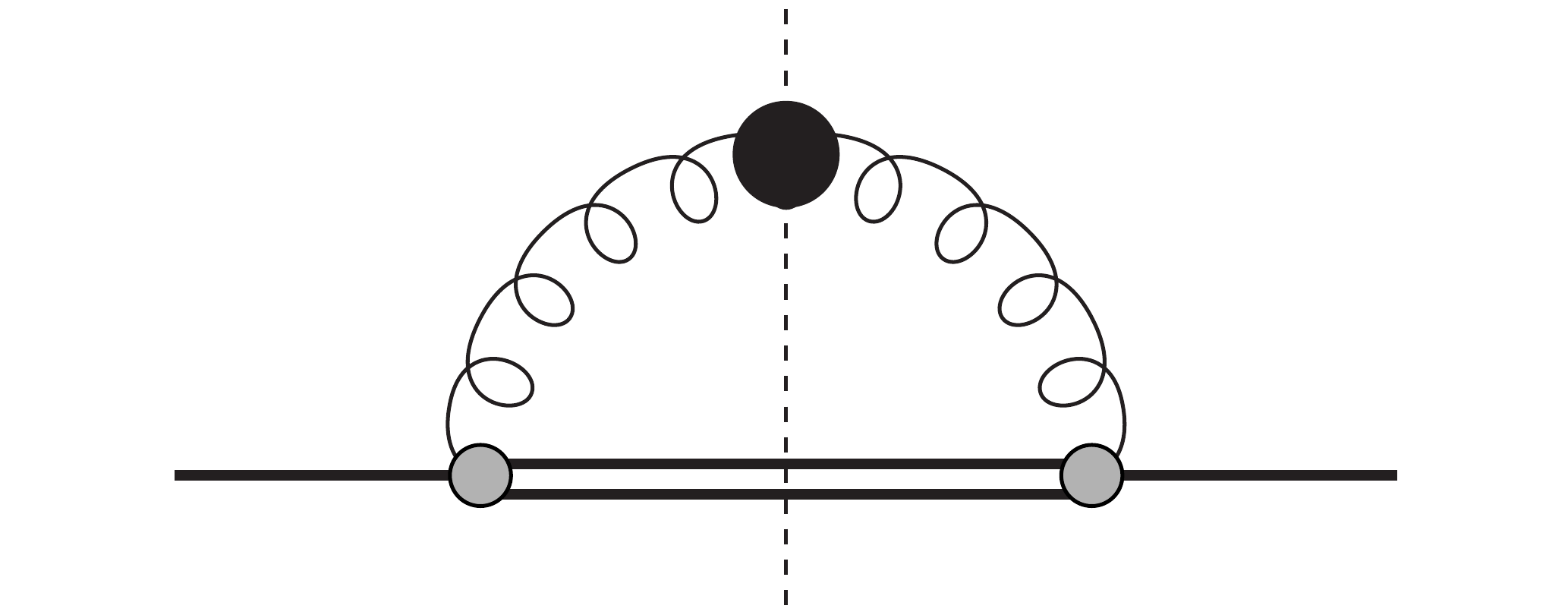}
\caption{ 
The self-energy diagram of the quarkonium color-singlet. 
At two-loop order, the cut diagram corresponds to the scattering processes of 
Fig. \ref{NLOdiag}. 
}
 \label{selfe}
\end{figure}

In comparison with the leading order result in Eq. (\ref{crossLO}), 
Eqs. (\ref{nlo-q}) and (\ref{nlo-g}) correspond to the dissociation 
cross sections at next-to-leading order. 
After multiplying by $d_p$, they agree with the pNRQCD results for the 
$mv\gg T\gg m_D\gg E$ regime where 
inelastic parton scattering is dominant in the 
effective field theory framework \cite{pnrqcd-nlo}. 
In pNRQCD, the NLO results are calculated by the 
imaginary part of the singlet potential which is from the 
self-energy diagram (see Fig. \ref{selfe}) of the quarkonium color-singlet. 
We note that the cut diagram at two-loop order corresponds to the scattering 
processes of Fig. \ref{NLOdiag}, and the leading contribution to 
gluo-dissociation of Fig. \ref{vertex}.

By convoluting the momentum distributions of incoming and outgoing partons, 
the thermal width by inelastic parton scattering is 
given by \cite{pnrqcd-nlo}
\begin{equation}
\Gamma_{NLO}=\int \frac{d^3\k_1}{(2\pi)^3} 
\bigg[d_q \, \sigma_{NLO}^{(q,\bar{q})}(k_1) \, 
n_F(k_1)[1-n_F(k_1)] +
d_g \, \sigma_{NLO}^{(g)}(k_1) \, 
n_B(k_1)[1+n_B(k_1)]\bigg] \, ,
\end{equation} 
where $n_{B}(k_1)$ and $n_{F}(k_1)$ are the Bose-Einstein and Fermi-Dirac 
momentum distributions, respectively. 
For hard ($\sim T$) momentum the thermal distribution of an initial parton is 
$n(k_{1})$ with $k_{10}\simeq k_1$, and  
the Bose-enhancement or Pauli-blocking factor of an outgoing parton 
has been approximated as $1\pm n(k_2)\simeq 1\pm n(k_1)$ for small energy 
transfer.

\section{Numerical Results for $\Upsilon(1S)$}
\label{numeric}

\begin{figure}
\includegraphics[width=0.45\textwidth]{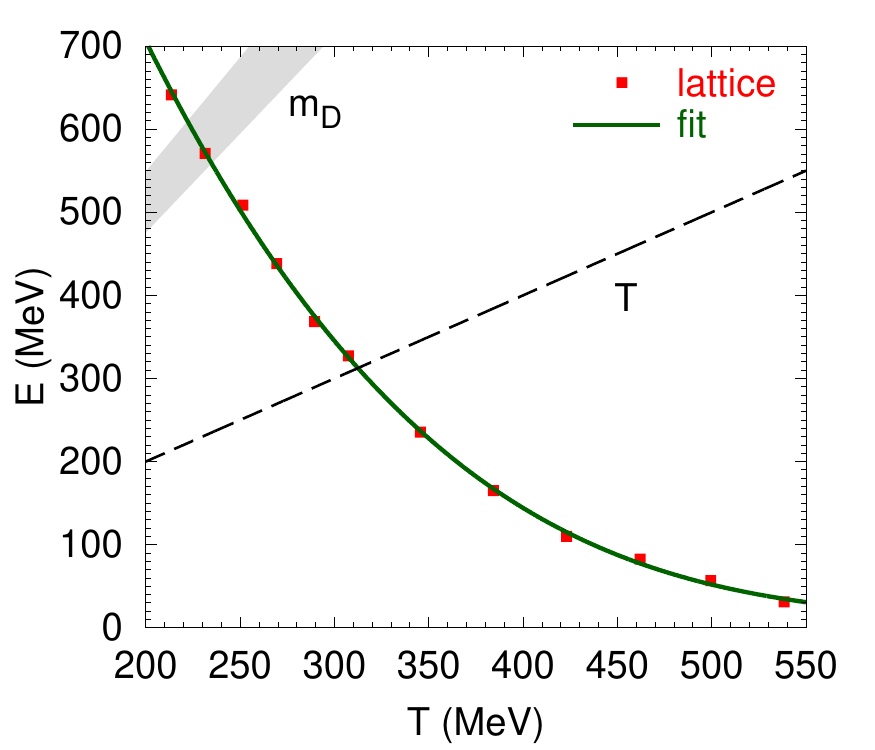}
\caption{
The binding energy of $\Upsilon(1S)$ from lattice computations \cite{lattice} 
and the numerical fit. 
The black dashed line shows the temperature for reference and the gray band is for the 
Debye mass as a function of temperature.  }
\label{lattice}
\end{figure}

In this section, we focus on the weakly bound $\Upsilon(1S)$ and present the 
numerical results of the dissociation cross sections and thermal widths. 
Above the phase transition temperature, the ground state of bottomonium 
is expected to survive up to $T\sim600$ MeV and melts at higher temperature 
\cite{strickland-review}. 
The binding energy of a Coulombic state depends on the coupling constant, 
$E\sim mg^4$ which increases as temperature decreases.  
To study transitional behaviors at the 
temperature regime $T\sim 200-500$ MeV, we set the effective coupling constant 
$\alpha_s=0.3-0.4$ and use the temperature-dependent binding energy which has 
been estimated in lattice QCD \cite{lattice}. 
The Debye screening mass given by 
$m_D^2=\frac{g^2T^2}{3}\left(N_c+\frac{N_f}{2}\right)$ is reduced with time as 
the temperature decreases with the evolution of quark-gluon plasma after  
heavy ion collisions.   
Although our analysis is based on a Coulombic bound state, 
the binding energy estimated from nonperturbative lattice 
computations might allow potential effects which are compatible with lattice 
QCD findings at short distances. 
Employing the temperature-dependent binding energy and screening mass, we 
simulate the running coupling which results 
$m_D\ll \pi T$ ($m_D\sim \pi T$) for weak (strong) coupling. 
With $N_c=N_f=3$, $m=4.8$ GeV and $a_0=0.14-0.18$ fm are used for 
$\Upsilon(1S)$.

Fig. \ref{lattice} shows an upper limit of the binding energy of 
$\Upsilon(1S)$ computed in lattice QCD \cite{lattice} and its numerical fit. 
While the Coulombic binding energy is determined as $243-432$ MeV for fixed 
coupling $\alpha_s=0.3-0.4$, the binding energy 
estimated from the lattice QCD decreases as temperature increases. 
The black dashed line represents the temperature ($E=T$) line, and 
the gray band is for the Debye mass as a function of $T$.

\begin{figure}
\includegraphics[width=0.45\textwidth]{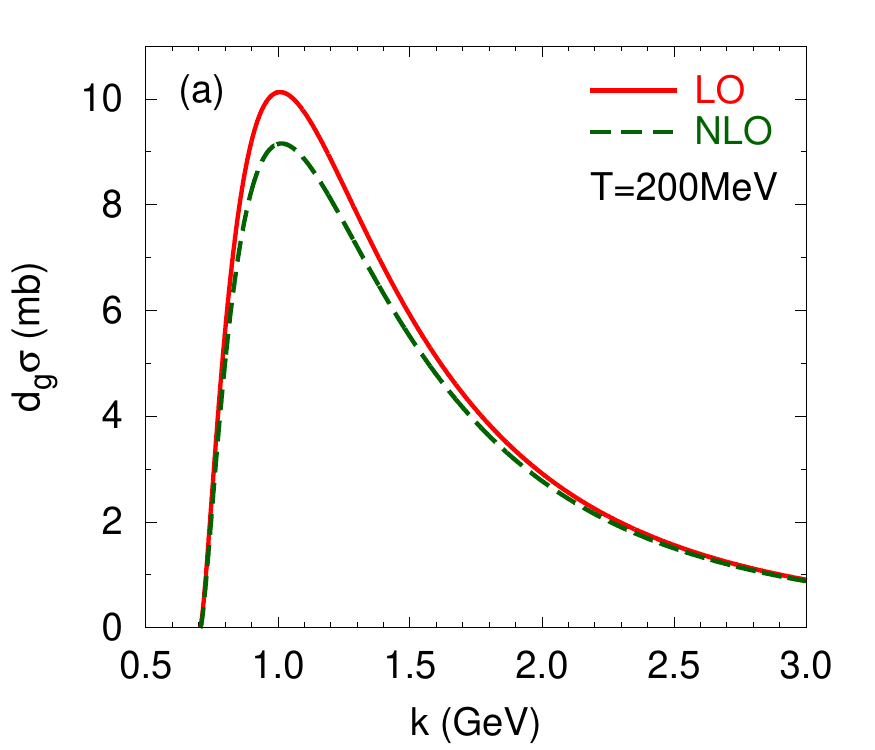}
\includegraphics[width=0.45\textwidth]{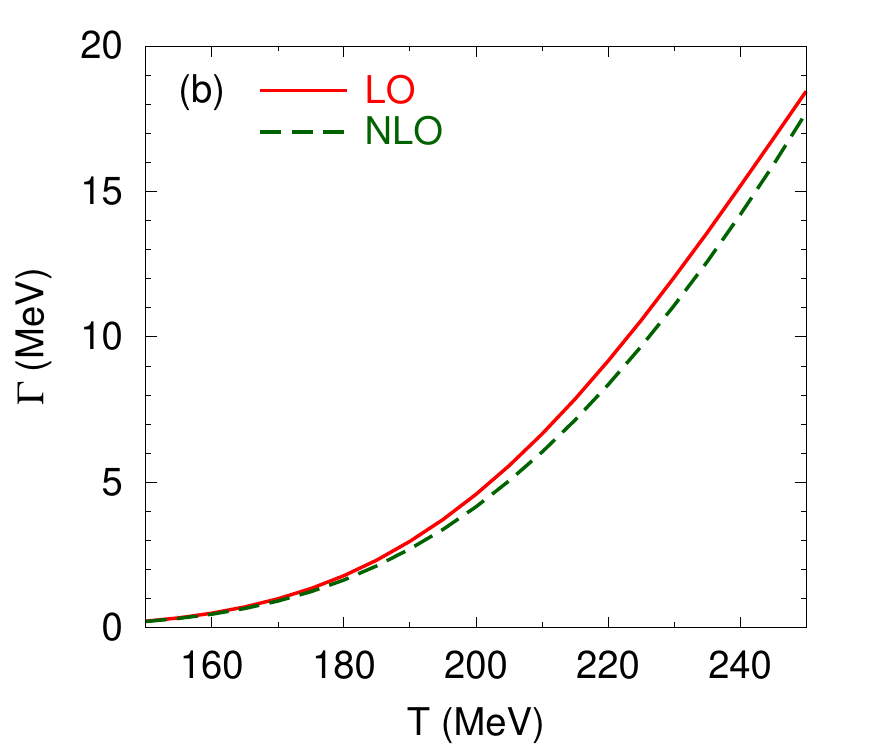}
\caption{
The leading order and next-to-leading order results by 
gluo-dissociation ($g+\Upsilon(1S)\rightarrow b+\bar{b}$) for $\alpha_s=0.4$. 
(a) $\Upsilon(1S)$ dissociation cross sections and 
(b) thermal widths. 
}
\label{gluonlo}
\end{figure}

Gluo-dissociation is the main mechanism for quarkonium 
dissolution at low temperature near the phase transition where the binding 
energy is relatively large. 
When an incoming gluon carries low momentum, gluo-dissociation contributes 
up to next-to-leading order. 
In Fig. \ref{gluonlo}, we present the dissociation cross sections and the 
thermal widths at leading (Eq. (\ref{crossLO})) and 
next-to-leading (Eq. (\ref{gluo-nlo})) orders. 
The next-to-leading order results are slightly smaller than the leading ones.  
Since the difference is relatively insignificant compared with that from 
inelastic parton scattering, we consider only leading order 
for gluo-dissociation in the following.

\begin{figure}
\includegraphics[width=0.45\textwidth]{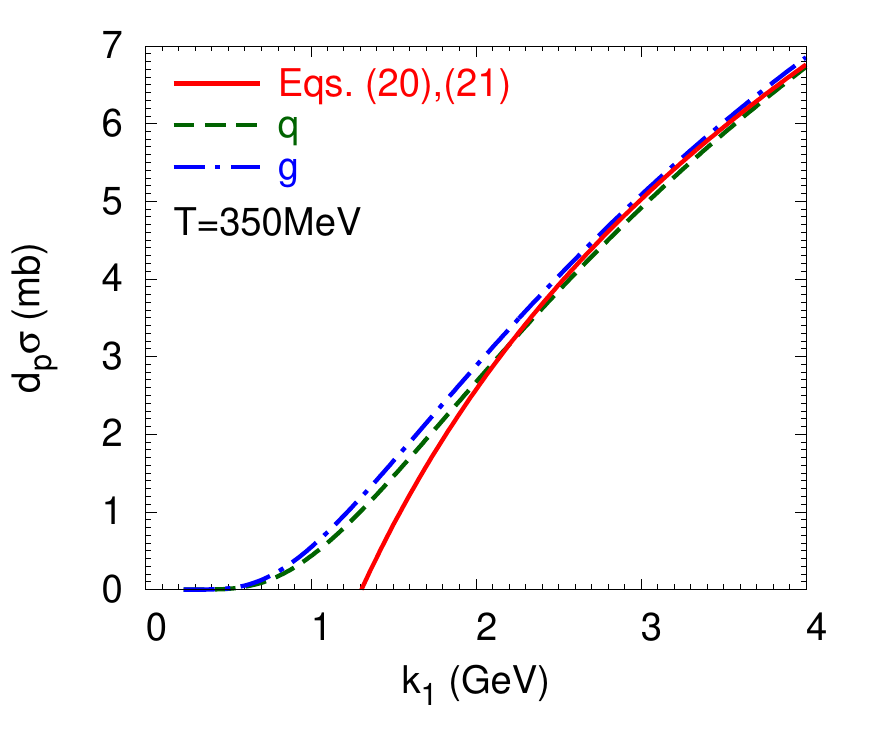}
\caption{
The comparison of the numerically calculated cross section with the 
asymptotic formula for $p+\Upsilon(1S)\rightarrow p+b+\bar{b}$ with 
$\alpha_s=0.4$. 
}
\label{numericsig}
\end{figure}

Inelastic parton scattering is important at the high temperature regime 
where the binding energy is smaller than $T$ \cite{rapp-review}. 
At high temperature, the Debye mass is large and the logarithmic formula in 
Eqs. (\ref{nlo-q}) and (\ref{nlo-g}) are not appropriate to study 
quarkonium dissociation. 
The cross sections become negative when an incoming parton has 
low momentum, 
which does not make sense (see the red solid line in Fig. \ref{numericsig}). 
They are also independent of the binding energy. 
These are because many restrictions are required to yield the logarithmic 
formula in the previous section. 
To obtain phenomenologically acceptable cross sections that can be used 
throughout a wide temperature region, we integrate the matrix elements 
of Eq. (\ref{mtx2-nlo}) over the phase space numerically. 
Since the matrix elements squared are functions of $k_1$, $k_2$, and the angle 
($\theta_{k_1k_2}$) between $\k_1$ and $\k_2$, integrating over the energy 
transfer ($k_0=k_1-k_2$) and $\cos\theta_{k_1k_2}$ yields the cross section 
as a function of $k_1$,   
\begin{equation}
\sigma_{NLO}(k_1)=\frac{1}{2^5(2\pi)^3d_\Upsilon d_p m_\Upsilon mk_1}
\int_E^{k_1} dk_0 \, (k_1-k_0)\int d(\cos\theta_{k_1k_2}) \, p|\mathcal{M}|_{NLO}^2\bigg\vert_{p^2=m(k_0-E)} \, .
\end{equation}
The numerically calculated cross sections are shown in Fig. \ref{numericsig} 
and compared with the analytical result. 
The cross sections vary from zero at $k_1=E$ and 
smoothly approach the asymptotic formula at high momentum.   
This method improves the cross sections in the near-threshold region.

\begin{figure}
\includegraphics[width=0.45\textwidth]{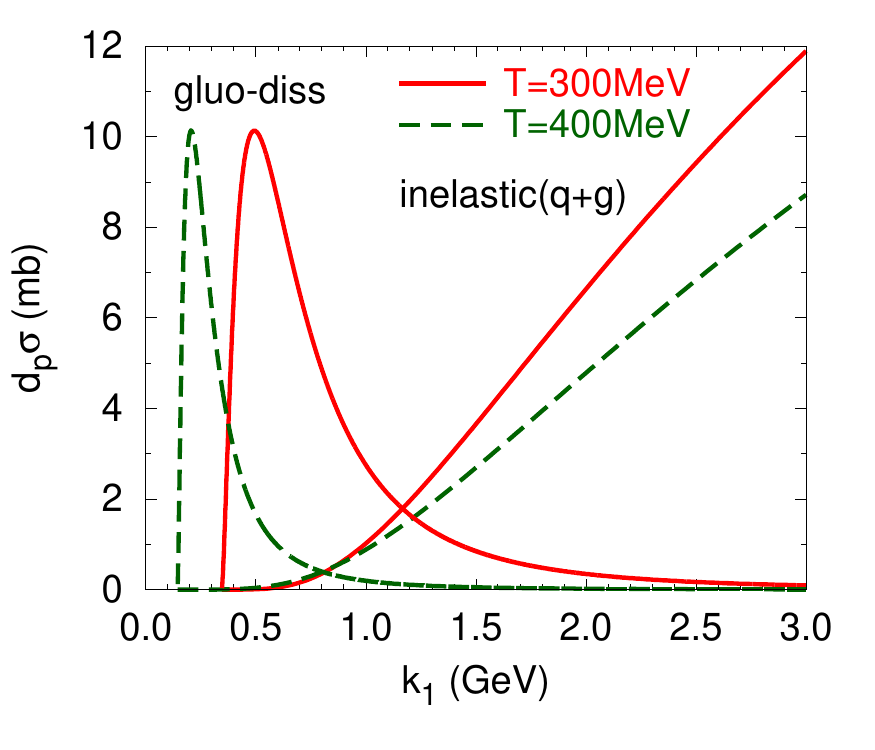}
\caption{
$\Upsilon(1S)$ dissociation cross sections by gluo-dissociation and 
inelastic parton scattering for $T=300,\, 400$ MeV and $\alpha_s=0.4$. 
}
\label{sigma2}
\end{figure}

In Fig. \ref{sigma2}, we present the cross sections of two mechanisms as 
a function of incoming parton momentum for two different temperatures. 
Gluo-dissociation is dominant at low momentum, 
while inelastic parton scattering is important when an incoming parton 
is energetic. 
The maximum of the gluo-dissociation cross section depends on the binding 
energy, and it shifts to lower momentum as temperature increases. 
In inelastic parton scattering, the cross section is smaller for higher 
temperature because the Debye screening mass increases with $T$.

\begin{figure}
\includegraphics[width=0.45\textwidth]{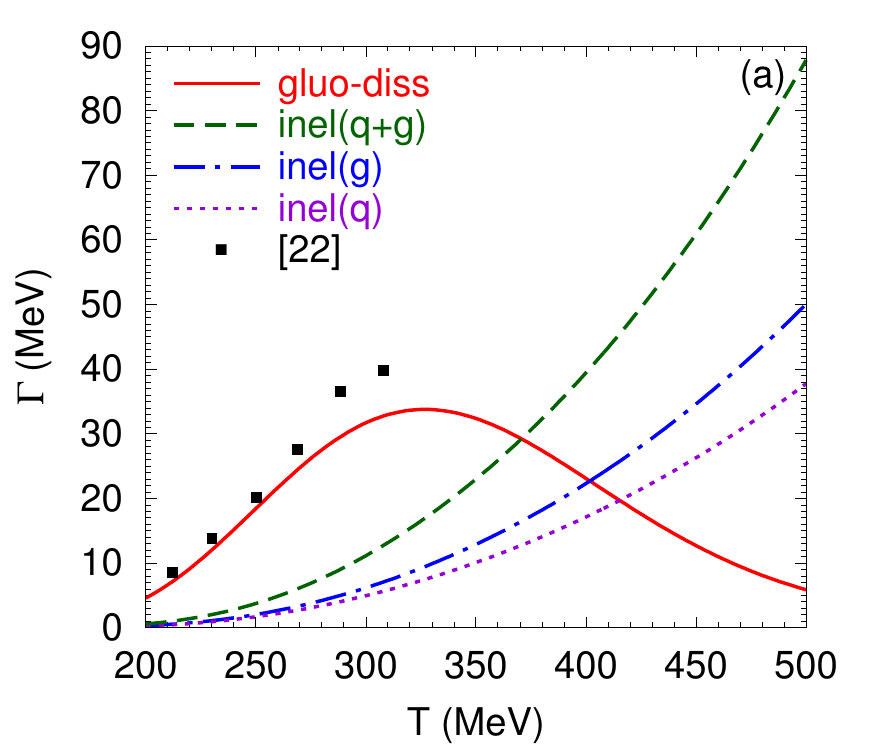}
\includegraphics[width=0.45\textwidth]{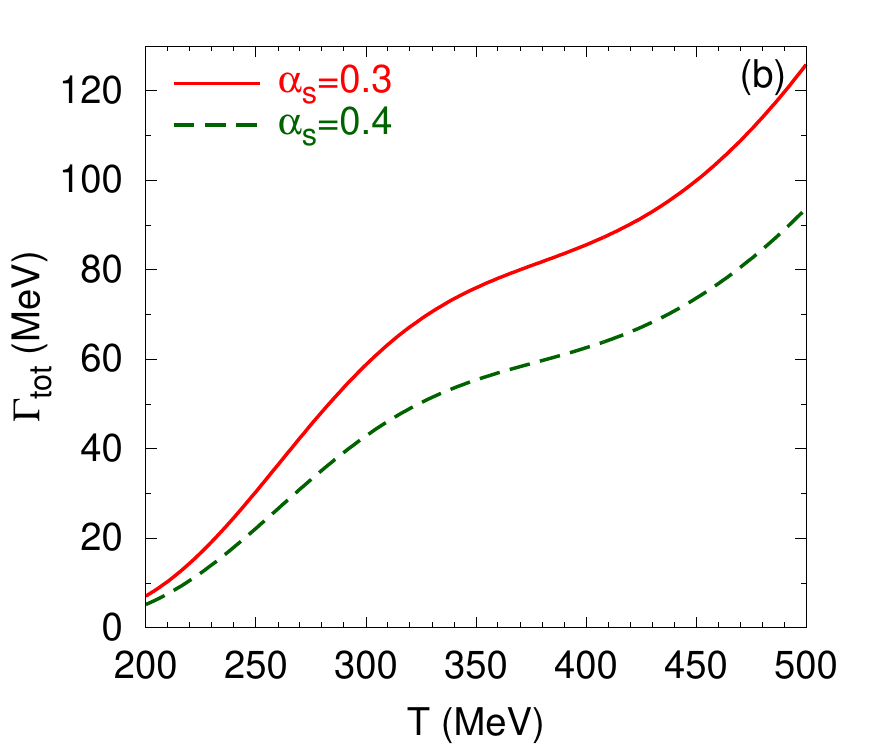}
\caption{
(a) Temperature dependence of the $\Upsilon(1S)$ thermal widths by 
gluo-dissociation and inelastic parton scattering for $\alpha_s=0.4$. 
(b) The total thermal width as the sum of the contributions by 
two mechanisms for $\alpha_s=0.3, \, 0.4$. 
}
\label{width}
\end{figure}

Fig. \ref{width} (a) shows the temperature dependence of the thermal width for 
each mechanism. 
As temperature decreases during the time evolution, the cross section of 
gluo-dissociation peaks at higher momentum. 
In the meanwhile, the phase space distribution $k^2 \, n_B(k)$ becomes 
smaller but peaks at lower momentum close to the maximum of the cross 
section. 
As a result, the thermal width by gluo-dissociation increases with time  
initially but decreases later at low temperature near the phase 
transition. 
This behavior seems to be contrasted with the leading result of Ref. 
\cite{pnrqcd-lo} in which the thermal width is proportional to $T$, but 
the width in pNRQCD is calculated with the Coulombic binding energy much 
smaller than the temperature scale which is different from the binding energy 
of Fig. \ref{lattice}.   
On the other hand, the thermal width by inelastic parton scattering decreases 
as a collision system cools down, as expected in Ref. \cite{width}.  
The reason is that although the cross section at lower temperature is larger, 
the phase space distribution $k_1^2 \, n(k_1)[1\pm n(k_1)]$ becomes 
much smaller (and peaks at lower momentum for inelastic quark scattering).  

Our numerical results indicate that the dominant mechanism of quarkonium 
dissociation changes from inelastic parton scattering at high temperature 
(where $E<T$) to gluo-dissociation near the phase transition (where $E>T$). 
These transitional behaviors depending on temperature are consistent with 
earlier analyses performed in the quasifree approximation \cite{rapp,rapp-review}. 
The black squares in Fig. \ref{width} (a) represent the width 
(extracted from Ref. \cite{lattice}) which is estimated nonperturbatively by 
tunnelling and direct thermal activation to the continuum in the limit of 
$E\gg T$ \cite{kharzeev}.  
They are almost same as the width by gluo-dissociation near the phase 
transition and comparable with the sum of those by two mechanisms at higher 
temperature.

Fig. \ref{width} (b) presents the total thermal width as the sum of the 
contributions by gluo-dissociation and inelastic parton scattering. 
Because the temperature dependence of inelastic parton scattering is stronger 
than that of gluo-dissociation, the total width increases with $T$. 
At high enough temperature where the thermal width exceeds the binding energy 
($\Gamma_{tot}\gtrsim E$), 
a heavy quark-antiquark pair is more likely to decay than 
to be bound so dissociation is expected \cite{strickland-review,lattice}. 
The weaker the coupling, the larger width we obtain because a smaller 
screening mass (larger Bohr radius) allows dissociation by inelastic 
parton scattering (gluo-dissociation) more probable. 
If we fix the screening mass as $m_D\sim 600$ MeV, the thermal width of 
inelastic parton scattering increases more rapidly with $T$. 
We have checked that variations of parameters do not change significantly 
the qualitative behaviors of the cross sections and thermal widths.  
Our results fairly agree with the old calculations in Ref. \cite{lee-width}.

\section{Higher order corrections}
\label{g-corr}

In Sec. \ref{inelastic}, we have obtained the next-to-leading order 
contributions to quarkonium dissociation by using the effective vertex derived 
from the leading order processes. 
If we ignore corrections to the effective vertex due to the heavy 
quark-antiquark interactions, higher order expansions might be possible. 
In the perspective of hard thermal loop (HTL) perturbation theory, the external 
partons are hard ($K_1,K_2\sim T$) and the exchange momentum is soft 
($K\sim gT$) in inelastic parton scattering. 
If all the gluons in Fig. \ref{NLOdiag} (b) are soft, $O(g)$ corrections arise 
because of the Bose enhancement effects \cite{soft}.

\begin{figure}
\includegraphics[width=0.4\textwidth]{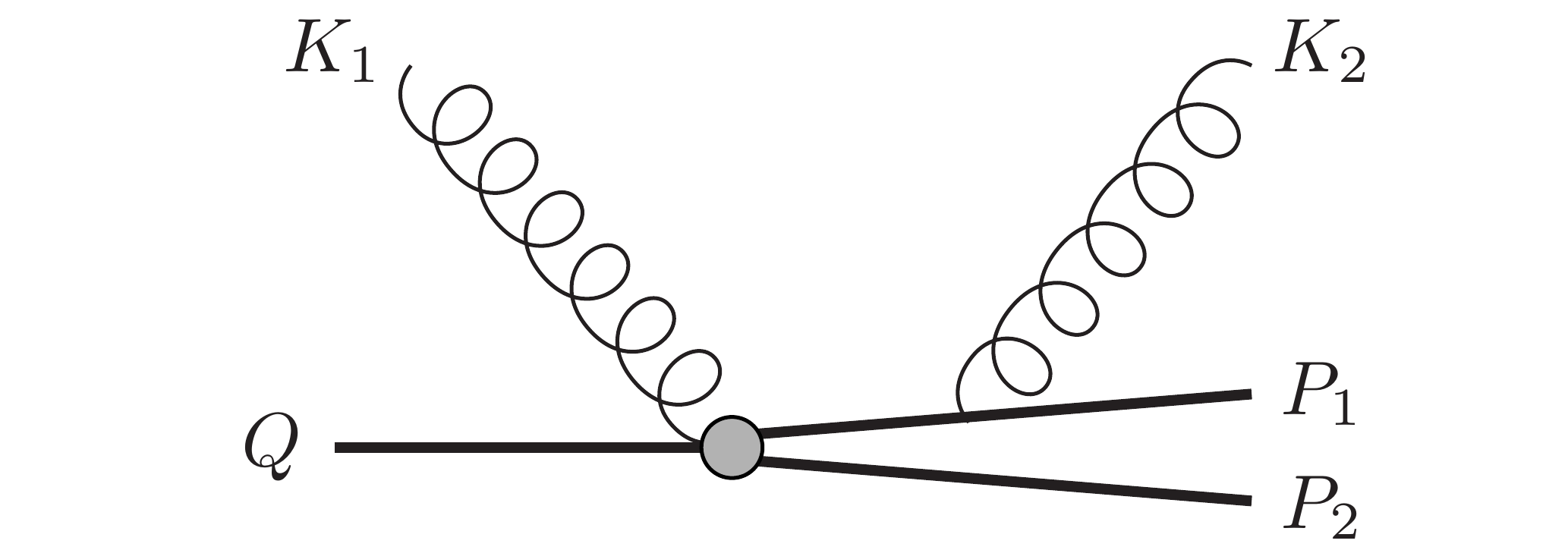}
\caption{
With soft gluons, $O(g)$ corrections to the next-to-leading order quarkonium 
dissociation can be obtained.  
}
 \label{NNLO}
\end{figure}

Another source of $O(g)$ corrections is the diagram in Fig. \ref{NNLO} (and 
the same except $P_1,P_2$ exchanged) with soft gluons. 
The scattering amplitude is given by \cite{lee-nlo}   
\begin{equation}
\label{mtx-O(g)}
\mathcal{M}_{O(g)}^{\mu\nu\lambda}= 
g\frac{g^{\lambda 0}}{k_{20}}\bar{u}(P_1)
V_0^{\nu\mu} (K_1)[T^a,T^b]v(P_2) \, ,
\end{equation}
where $V_0^{\nu\mu}(K_1)$ is same as $V^{\nu\mu}(K_1)$ without $T^a$. 
As in Sec. \ref{inelastic}, only the longitudinal gluon part of the 
effective vertex contributes. 
By comparing Eqs. (\ref{mtx-nlo}) and (\ref{mtx-O(g)}), we note that 
\begin{equation}
\frac{|\mathcal{M}_{O(g)}|^2}{|\mathcal{M}_{NLO}^{(g)}|^2}\sim
\frac{k^2}{k_1^2} \, .
\end{equation}
When $K_1, K_2$ are hard, the process of Fig. \ref{NNLO} is suppressed, 
giving $O(g^2)$ corrections to the inelastic parton scattering of 
Fig. \ref{NLOdiag}. 
However, if the gluons are soft 
two matrix elements yield the same order of magnitude. 
These $O(g)$ corrections have been calculated to obtain the momentum 
diffusion coefficients of a heavy quark at next-to-leading order 
\cite{sch-moore1,sch-moore2}.

The momentum diffusion coefficient of a heavy quark is defined as the 
mean-squared momentum transfer per unit time. 
At leading order, it is given by \cite{moore-teaney} 
\begin{multline}
\label{kappa}
3\kappa_{LO}=\frac{1}{2m}
\int\frac{d^3\k_1}{(2\pi)^32k_{10}}
\int\frac{d^3\k_2}{(2\pi)^32k_{20}}
\int\frac{d^3\p_2}{(2\pi)^32p_{20}}
\\
\times 
(2\pi)^4\delta^4(K_1+P_1-K_2-P_2)
(\p_2-\p_1)^2|\mathcal{M}|_{LO}^{2(\kappa)}n(k_1)
[1\pm n(k_1)] \, , 
\end{multline}
where $\mathcal{M}_{LO}^{(\kappa)}$ is the matrix element 
for $K_1+P_1\rightarrow K_2+P_2$. 
The corresponding scattering processes are similar to those 
of Fig. \ref{NLOdiag}. 
Instead of barely bound quarkonium breaking into a heavy quark and 
antiquark, a nonrelativistic heavy quark scatters with bath particles 
exchanging soft momentum. 
The factor $(\p_2-\p_1)^2$ in Eq. (\ref{kappa}) corresponds to the momentum 
transfer $(\k_1-\k_2)^2$ in Eq. (\ref{mtx2-nlo}) which comes from the 
longitudinal gluon part of the effective vertex. 
By noting that the thermal width has one more momentum integration with a bound 
state, we obtain the following correspondence \cite{eft}: 
\begin{equation}
\Gamma_{NLO}
\simeq
\langle r^2\rangle\kappa_{LO} \, ,
\end{equation}
where 
$\langle r^2\rangle=\int \frac{d^3\p}{(2\pi)^3} |\nabla\psi(\p)|^2=3a_0^2$ 
for the Coulombic wave function of Eq. (\ref{coulomb}). 
In this regard, $\kappa_{NLO}$ in Ref. \cite{sch-moore1,sch-moore2} might be 
useful to obtain $O(g)$ corrections to the thermal width 
$\Gamma_{NLO}$ \cite{pnrqcd-nlo}. 
However, because there might be other corrections such as those induced by 
the heavy quark-antiquark interactions, we do not proceed further in this work.

\section{summary}
\label{summary}

We have introduced a partonic picture for quarkonium dissociation that 
reduces to the formal limit of effective field theory calculations 
at the relevant kinematical regime, but that can also be employed in a wide 
temperature range which is applicable to heavy quark systems in 
evolving plasma systems. 
In particular, 
we have discussed two mechanisms of quarkonium dissociation, 
gluo-dissociation and inelastic parton scattering, in the limit of small 
energy transfer which is suitable to study weakly bound quarkonia. 

Gluo-dissociation is related to the plasmon pole contribution, 
and inelastic parton scattering is induced by the Landau damping phenomenon. 
By using hard thermal loop perturbative theory, we have rederived the 
dissociation cross sections and calculated the thermal widths. 
The thermal effects at next-to-leading order are obtained through the 
gluon dispersion relation and resummed propagator 
for gluo-dissociation and inelastic parton scattering, respectively. 
Our results in Eqs. (\ref{gluo-nlo}), (\ref{nlo-q}), and (\ref{nlo-g}) 
agree with those obtained by 
potential nonrelativistic QCD in which the thermal width is calculated from the 
imaginary part of the singlet potential.  
This might imply that the imaginary parts of the heavy quark-antiquark 
potential responsible for quarkonium dissociation basically originate from 
various scattering processes in thermal medium, at least for weakly bound 
quarkonia. 
By comparing with the momentum diffusion coefficient of a heavy quark, we have 
discussed possible $O(g)$ corrections to the next-to-leading order quarkonium 
dissolution.

To study the transitional behaviors with the running coupling effects, 
we have employed the nonperturbative lattice input of binding energy in the 
analysis of $\Upsilon(1S)$ dissolution.  
As discussed in the quasifree approximation \cite{rapp,rapp-review}, 
gluo-dissociation is important at low temperature near the phase transition 
while inelastic parton scattering becomes dominant at high temperature when an 
incoming parton carries high momentum. 
Our numerical approach might be useful for phenomenological studies of 
quarkonia transport in relativistic heavy ion collisions.

\section*{Acknowledgments}
\begin{sloppypar}
This work is supported by the National Research Foundation of Korea
(NRF) grant funded by the Korea government (MSIT) 
(No. 2018R1C1B6008119 and  No. 2016R1D1A1B03930089).
\end{sloppypar}

\end{document}